\documentclass[useAMS,usenatbib]{mn2e}
\usepackage{graphicx}
\usepackage{latexsym}
\usepackage{amsmath}
\usepackage{url}

\title[Blueshifted OI lines from PPDs]{Blueshifted [OI] lines from protoplanetary discs: the smoking gun of X-ray photoevaporation.}

\author[Ercolano, Owen]{Barbara Ercolano$^{1,2}$\thanks{E-mail: ercolano@usm.lmu.de (BE)}, James E. Owen$^{3,4}$\\
$^{1}$Universit\"ats-Sternwarte M\"unchen, Scheinerstr. 1, 81679 M\"unchen, Germany\\
$^{2}$Excellence Cluster Origin and Structure of the Universe,
Boltzmannstr.2, 85748 Garching bei M\"unchen, Germany\\
$^3$Institute for Advanced Study, Einstein Drive, Princeton, NJ 08540, USA\\
$^4$Hubble Fellow}

\begin{document}

\pagerange{\pageref{firstpage}--\pageref{lastpage}} \pubyear{2011}

\maketitle

\label{firstpage}

\def\nat{Nature}
\def\mnras{MNRAS}
\def\apj{ApJ}
\def\aap{A\&A}
\def\apjl{ApJL}
\def\apjs{ApJS}
\def\bain{BAIN}
\def\araa{ARA\&A}
\def\pasp{PASP}
\def\aj{AJ}
\def\pasj{PASJ}
\def\ga{\sim}

\begin{abstract}

Photoevaporation of protoplanetary discs by high energy radiation from the central young stellar object is currently the favourite model to explain the sudden dispersal of discs from the inside out. While several theoretical works have provided a detailed pictured of this process, the direct observational validation is still lacking. 
Emission lines produced in these slow moving protoplanetary disc winds may bear the imprint of the wind structure and thus provide a potential diagnostic of the underlying dispersal process. 
In this paper we primarily focus on the collisionally excited neutral oxygen line at 6300A. We compare our models predictions to observational data and demonstrate a thermal origin for the observed blueshifted low-velocity component of this line from protoplanetary discs. Furthermore our models show that while this line is a clear tell-tale-sign of a warm, quasi-neutral disc wind, typical of X-ray photoevaporation, its strong temperature dependence makes it unsuitable to measure detailed wind quantities like mass-loss-rate.
\end{abstract}


\section{Introduction}

Understanding disc dispersal is a key piece in the puzzle of planet
formation as it sets the timescale over which
(gas giant) planet formation must occur. Furthermore the similarity
between the observed timescales over which Young Stellar Objects
(YSOs) lose their disc and the theoretically estimated timescale for
the formation of planets suggests that the two processes are probably
coupled and feed back on each other. The final build up of gaseous
planets occurs in discs on their last gasp before dispersal, and the dispersal process itself can lead to rapid evolution of young planets (Owen \& Wu, 2016), 
making evolved discs all the more interesting to study.  

A popular model to drive disc dispersal is
photoevaporation by radiation from the central star (e.g. Clarke et
al. 2001). The exact nature of the driving
radiation is however still open to debate.  (Extreme and Far)
Ultraviolet (UV) radiation as well as X-ray radiation has been shown to be able
to drive winds from the disc upper layers (Alexander et al. 2006;
Gorti, Hollenbach \& Dullemond 2009; Ercolano et al. 2008, 2009; Owen et
al. 2010) that are efficient enough to disperse the discs in the observed
timescales. However both the location and intensity of the wind depend
strongly on the driving radiation, with differences of more than two
orders of magnitude for mass loss rates predicted by different
models. Thus these different models obviously have profound implications for disc evolution and hence
for the formation of planets and their subsequent evolution
(e.g. Ercolano \& Rosotti 2015). 

The presence of a warm, at least partially, ionised disc wind has been confirmed via the
observation of a few km/s blue-shift in the profile of the
[NeII]~12.8$\mu$m fine structure line (Pascucci et
al. 2007). Unfortunately  modelling of
this line cannot shed light on the radiation source which drives the wind, due to
the different routes to formation of the Ne$^+$ ion which have different efficiencies in a fully ionised
EUV-driven wind and in a quasi-neutral X-ray driven wind. Ne$^+$
formation occurs via the removal of a valence electron from Ne atoms
in the fully-ionised winds driven by EUV radiation. In an X-ray driven wind Ne$^+$ is predominantly produced via charge exchange of Ne$^{2+}$ with neutral H atoms, which are abundant in the
quasi-neutral winds driven by X-rays. Atomic physics thus conspires to
the result that both an EUV- and an X-ray driven
wind, whose mass-loss rates differ by over two orders of magnitude, can equally well fit the observations (Ercolano \&
Owen 2010; Alexander 2008; Pascucci et al. 2011). 

Forbidden lines from low-ionisation and atomic states of common
elements may present an alternative way 
to study the wind dispersal mechanism. In particular a few km/s
blueshift has been measured in the profiles of, for example, the [OI] 6300,
[O I] 5577, [S II] 6731, and [N II] 6583 lines  (e.g. Hartigan et al. 1995, White \& Hillenbrand 2004, Mohanty
et al. 2005, Rigliaco et al. 2013, Natta et al. 2014). The profiles are often
double-peaked, with one component blue-shifted to a few hundred km/s
(high velocity component, HVC) and a second component typically
blue-shifted by only a few km/s (low velocity component, LVC), typical
of a photoionised wind. The HVC is generally attributed to emission
from a dense outflows closer to the star or jets (Hartigan et al. 1995). The collisionally excited neutral oxygen line at 6300A, in particular, gained significant attention for its potential to discriminate between an EUV- and an X-ray driven wind. EUV-driven winds are, by construction, fully ionised and cannot match the observed luminosities of the LVCs of the [OI] 6300 line (Font et al. 2004). X-ray winds, on the other hand, are only weakly ionised and warm enough to produce the [OI] 6300 line. 

Ercolano \& Owen (2010, EO10)  calculated synthetic spectra from the
X-ray photoevaporation models of Owen et al. (2010,2011), providing an
atlas of atomic and low-ionising emission lines, including the [OI] 6300 line. The line intensities
and profiles predicted by EO10 were in agreement with the observations
available at the time, thus suggesting that the blue-shifted LVC of the [OI] 6300 line is produced via collisional excitation of neutral oxygen atoms with electrons and hydrogen atoms in the slow-moving photoevaporative winds driven by X-ray radiation. 

Later observational studies by Rigliaco et
al. (2013) and Natta et al. (2014) showed some significant inconsistencies between the model predictions of the [OI] 6300 line by EO10 and the new data. This led Rigliaco et al. (2013) to suggest a non-thermal origin for the [OI] 6300 line (see also Gorti et al. 2011). Natta et al. (2014), on the other hand, still argue against a non-thermal origin of the OI lines, based on the simultaneous presence of the [SII] 4068 line in their spectra.

In this paper we revisit the theoretical models in light of these later observations and show that the [OI] 6300
data can be indeed explained in the context of an X-ray driven
photoevaporative wind and confirm a thermal origin for this line. We show that the
over-simplistic scaling of the illuminating flux in the work by EO10 is
to blame for the misleading conclusions on the origin of the [OI] 6300. 
In Section 2 we present the methods employed in our study. In Section 3 we describe the modelling strategy and present the results. A final discussion and our conclusions are given in Section 4.

\section{Methods}

\subsection{X-ray photoevaporative wind structure}
We use the set of wind solutions (density and velocity distribution of gas in the wind) for primordial discs 
(i.e. gas-rich, optically thick discs, which do not have an evacuated inner cavity) calculated by Owen et al. (2010,2011) and EO10 for a 0.7 M$_{\odot}$ star and X-ray luminosities (0.1 keV $\leq h\nu \leq 10$ keV) of $L_X = 2\times 10^{28},  2\times 10^{29}$ and  $2\times 10^{30}$ erg/sec. These were obtained by means of two-dimensional hydrodynamic calculations using the {\sc zeus} code (Stone et al. 1992a,b,c; Hayes et al. 2006), modified to include the effects of X-ray irradiation with a parametrisation of the gas temperature as a function of the local ionisation parameter. The dust radiative transfer and photosionisation code {\sc mocassin} (Ercolano et al. 2003, 2005, 2008b), modified according to Ercolano et al (2008a), was used produce the temperature parametrisation. The atomic database of the {\sc mocassin} code included opacity data from Verner et al. (1993) and Verner \& Yakovlev (1995), energy levels, collision strengths and transition probabilities from Version 5.2 of the CHIANTI database (Landi et al. 2006, and references therein) and hydrogen and helium free-bound continuous emission data of Ercolano \& Storey (2006). The ionising spectrum used to calculate the temperature parametrisation was calculated by Ercolano et al (2009a), using the plasma code of Kashyap \& Drake (2000) from an emission measure distribution based on that derived for RS CVn type binaries by Sanz-Forcada et al. (2002), which peaks at 10$^4$ K and fits to Chandra spectra of T-Tauri stars by Maggio et al. (2007), which peaks at around 10$^{7.5}$ K. This spectrum has a significant EUV component (13.6 eV $\leq h\nu \leq 0.1 keV$), with roughly  L$_{EUV}$ = L$_X$.  Solar abundances (Asplund et al. 2005), depleted according to Savage \& Sembach (1996) were assumed, namely (number density, with respect to hydrogen): $He/H = 0.1, C/H =
1.4\times10^{−4}, N/H = 8.32 \times 10^{−5}, O/H = 3.2 \times 10^{−4}, Ne/H = 1.2 \times 10^{−4}, Mg/H = 1.1 \times 10^{−6}, Si/H = 1.7 \times 10^{−6}, S/H = 2.8 \times 10^{−5}$. More details about the codes and setup of the models can be found in Ercolano et al. (2008a, 2009a) and Owen et al. (2010). 

The hydrodynamical calculations where performed in spherical co-ordinates with a domain spanning $[0,\pi/2]$ in the $\theta$ direction and $[r_{\rm in},r_{\rm out}]$ in the radial direction, with $r_{\rm out}$ set to 100~AU. In the calculations taken from Owen et al. (2010,2011) $r_{\rm in}$ was set to 0.33~AU and a resolution of 100 uniformly spaced cells in the angular direction and 250 non-uniformly spaced cells in the radial was used. As we shall discuss later in order to asses the role of [OI] emission from the bound inner disc ($R<1$~AU) we perform a new set of hydrodynamical calculations, this time with $r_{\rm in}$=0.04~AU with a resolution of 256 uniformly spaced cells in the angular direction and 384 non-uniformly spaced cells in the radial direction. Since the wind is launched from approximately 1~AU in the calculations, a smaller inner boundary did not effect the dynamics it just gave the hydrostatic density and temperature structure of the inner disc which we could use to calculate the [OI] emissivities. In all cases the simulations were run for at least 10 dynamical time-scales at the outer boundary until a steady-state was achieved (see discussion in Owen et al. 2010). 

\subsection{Photoionisation calculations}
Following the approach in EO10, we perform photoionisation calculations of the wind structures with the aim of predicting the intensity and spectral profile of the collisionally excited neutral hydrogen line at 6300A and compare it with the observational results of Rigliaco et al. (2013). The {\sc mocassin} code is  employed for this task, with exactly the same settings as described in the previous section. The ionising spectrum we use is the same as described above for the X-ray region, and used by EO10, additionally we consider here a softer spectral component due to accretion on the young stellar object (YSO). The latter is approximated as a blackbody of temperature 12000K and luminosity expressed as a multiple of the stellar bolometric luminosity, L$_{acc}$ = a $\times$ L$_{bol}$, with $a$ ranging from 10 to 10$^{-5}$. It is important to note that since the wind itself is driven by the X-rays (in the range 0.1-1~keV), modifying the soft UV spectrum will not effect the dynamics of the wind itself (Owen et al. 2012), and it is thus not necessary to run new wind simulations for the purpose of this work. 

\subsection{Emissivity and line profile calculations}
We have used our two-dimensional map of emissivities and gas
velocities obtained from {\sc mocassin} to reconstruct a three-dimensional cube of the disc and
calculate the line-of-sight emission profiles for several of the [OI] transitions. The emissivities are calculated using 120 logarithmically spaced radial points ($N_R$=120)\footnote{Note throughout this paper we use $\{r,\theta,\phi\}$ and $\{R,\varphi,z\}$ to distinguish between spherical and cylindrical polar co-ordinates respectfully.}  and 1500 logarithmically spaced height points (N$_Z$=1500). By assuming azimuthal and reflection symmetry the resultant 3D grid has dimensions $N_R \times 2N_Z \times N_\varphi$, where we adopt a values of $N_\varphi=400$ which was sufficient to resolve the line profiles. For the [OI] lines the disc atmosphere is optically thin  and the contribution to the line can escape freely, provided that the line of sight does not intercept the disc mid-plane which is completely optically thick due to dust (c.f. EC10). We neglect attenuation due to dust in the disc's atmosphere, this will only effect the results for the largest inclinations. 

The line luminosity is then computed by including a Doppler broadening term in each cell. Thus, the luminosity at a given velocity $u$ is computed using numerical integration by direct summation, taking the emissivities and velocity to be constant in each cell. Such that the line luminosity at a given velocity $L(u)$ is given by:
\begin{equation}
L(u)=\int\!\!\!{\rm d}^3r\frac{\ell(r)}{\sqrt{2\pi v_{\rm th}(r)^2}} \dot \exp \left(-\frac{\left[u-u_{\rm los}(r) \right]^2}{2v_{\rm th}(r)^2} \right)
\end{equation}
where $\ell(r)$ is the volume averaged power emitted at a point $r$, $u_{\rm los}$ is the projected gas velocity along the line of sight and $v_{\rm th}$ is the local rms velocity of the emitting atom. The lines were computed with a velocity resolution of 0.25~km~s$^{-1}$. The lines were then degraded to an instrumental resolution of $R=25,000$ -- i.e. the resolution of the Hartigan et al. (1995) and Rigliaco et al. (2013) study -- and $R=50,000$ a higher resolution representative of a future study. The degradation was performed by convolving the line profiles with a Gaussian profile of the appropriate width. The line profiles were calculated for disc inclinations of 0 to 90 degrees at 10 degree intervals.   

\section{Strategy and Results}
On the basis of their observations Rigliaco et al. (2013) highlighted a number of discrepancies with the models of EO10, which argued against a thermal origin of the [OI] 6300 line in an X-ray driven photoevaporative wind, as suggested by EO10. In particular, in contrast to the models of EO10, the observations showed: (1) no correlation between the [OI] luminosity, L$_{[OI]}$, and the X-ray luminosity, L$_X$, (2) a correlation of L$_{[OI]}$ with the FUV luminosity L$_{FUV}$, (3) higher full width half maximum (FWHM) of the [OI]6300 line and (4) a lower [OI]6300/[OI]5577 ratio.

We will show here that the above discrepancies can be fully explained by considering the assumptions made by EO10, with regards to the ionising spectrum of the central Young Stellar Object (YSO). As described above, the illuminating spectrum used by EO10 extends to the EUV region, with L$_{EUV}$ = L$_X$. In order to explore the relation between of L$_{[OI]}$ and L$_X$, EO10 scaled their {\it entire} spectrum by the same factor, hence also increasing/decreasing the L$_{EUV}$ reaching the X-ray driven wind. We suggest here that the L$_{[OI]}$-L$_X$ correlation reported by EO10 is actually the L$_{[OI]}$-L$_{EUV}$ correlation, resulting from the homogeneous scaling of the input spectrum at all wavelengths. We demonstrate this here by performing the numerical experiment of decoupling the UV and the X-ray regions of the illuminating spectrum, where we define X-ray/EUV energies higher/lower than 0.1 keV, respectively. Practically we include an additional input spectrum in the form of a blackbody of temperature 12000K, loosely representative of an accretion component, whose luminosity can be scaled independently from the X-ray luminosity. We then set up a grid of models at constant  L$_{acc}$ and vary only L$_X$ and a grid of models at constant L$_X$ and vary only  L$_{acc}$. The models are summarised in Table 1 and the results are described in the next section. 

For the figures presented in this section, we used all the observational data reported in the Rigliaco et al. (2013) study, which was partially based on re-analysis of previous data by Hartigan et al. (1995).

\begin{figure}
\begin{center}
\includegraphics[width=0.47\textwidth]{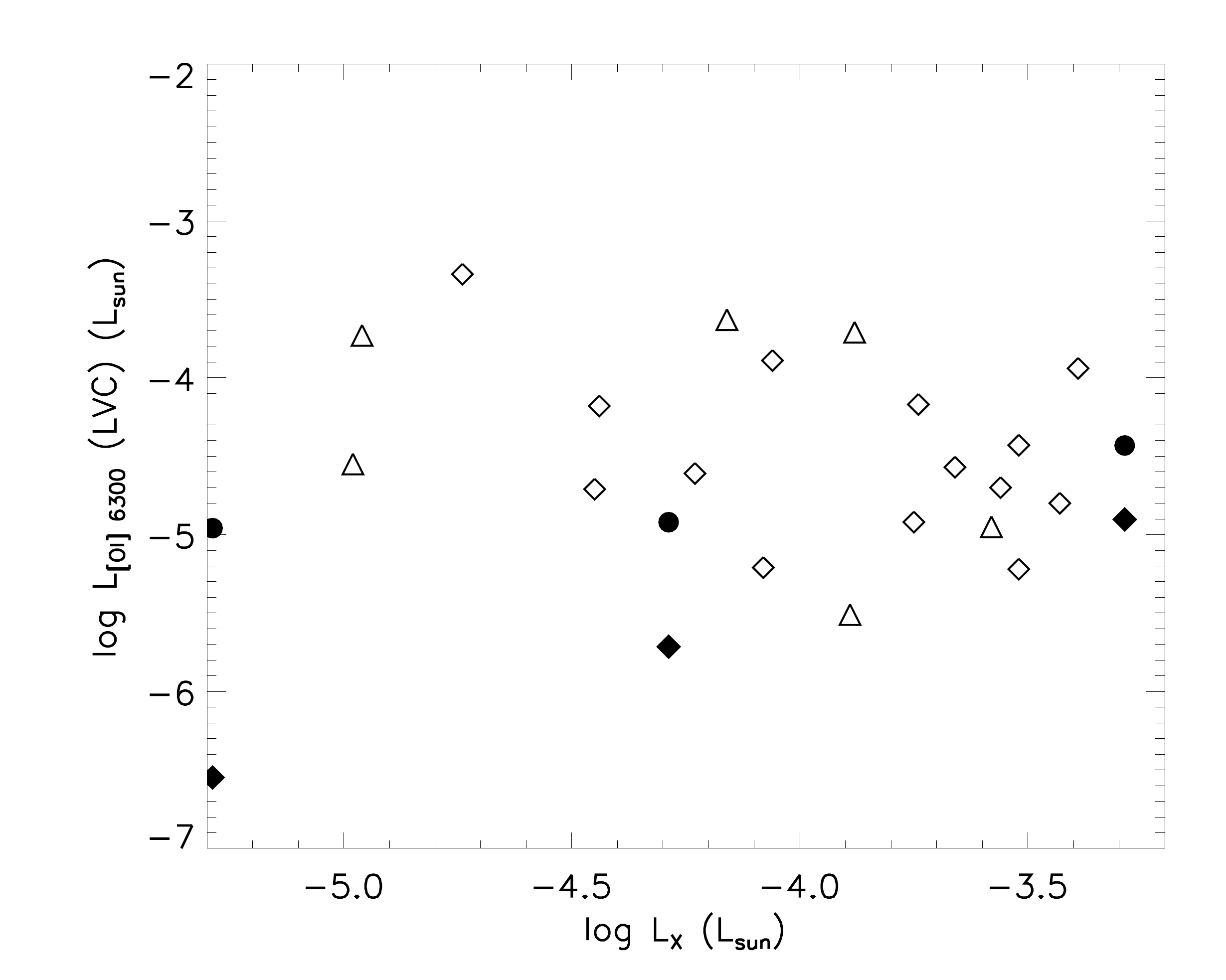}
\caption{[O I] 6300 luminosity versus X-ray luminosity for the
subsample of 21 Sample II objects from Rigliaco et al. (2013) (empty diamonds and triangles, where triangles denote upper limits in X-ray luminosity), compared to the model predictions of EO10 (filled diamonds) and those from this work with constant accretion luminosity and suppressed chromospheric emission $<$100eV  (filled circles). See section 3.1 for details.}
\end{center}
\end{figure}

\subsection{Models at constant accretion luminosity: no correlation with L$_X$}
EO10 pubilshed [OI]6300 line intensities obtained from their X-ray photoevaporative wind model of a 0.7~M$_{\odot}$ star with L$_X  =  2\times 10^{28},  2\times 10^{29}$ and  $2\times 10^{30}$ erg/sec. The resulting L$_{[OI]}$ showed a near linear correlation with L$_X$ (filled diamonds in Figure 1), which is not seen in the observations of Rigliaco et al. (2013) (empty symbols in Figure 1). As mentioned already in the previous section, we show here that this apparent correlation is driven by the L$_{EUV}$-L$_X$ correlation in the scaling of the illuminating spectra in the EO10 model. Indeed, as shown by the filled circles in Figure 1, our models with constant accretion luminosity where we vary L$_X$ only do not show any correlation at all between L$_{[OI]}$-L$_X$.

\begin{figure}
\begin{center}
\includegraphics[width=0.47\textwidth]{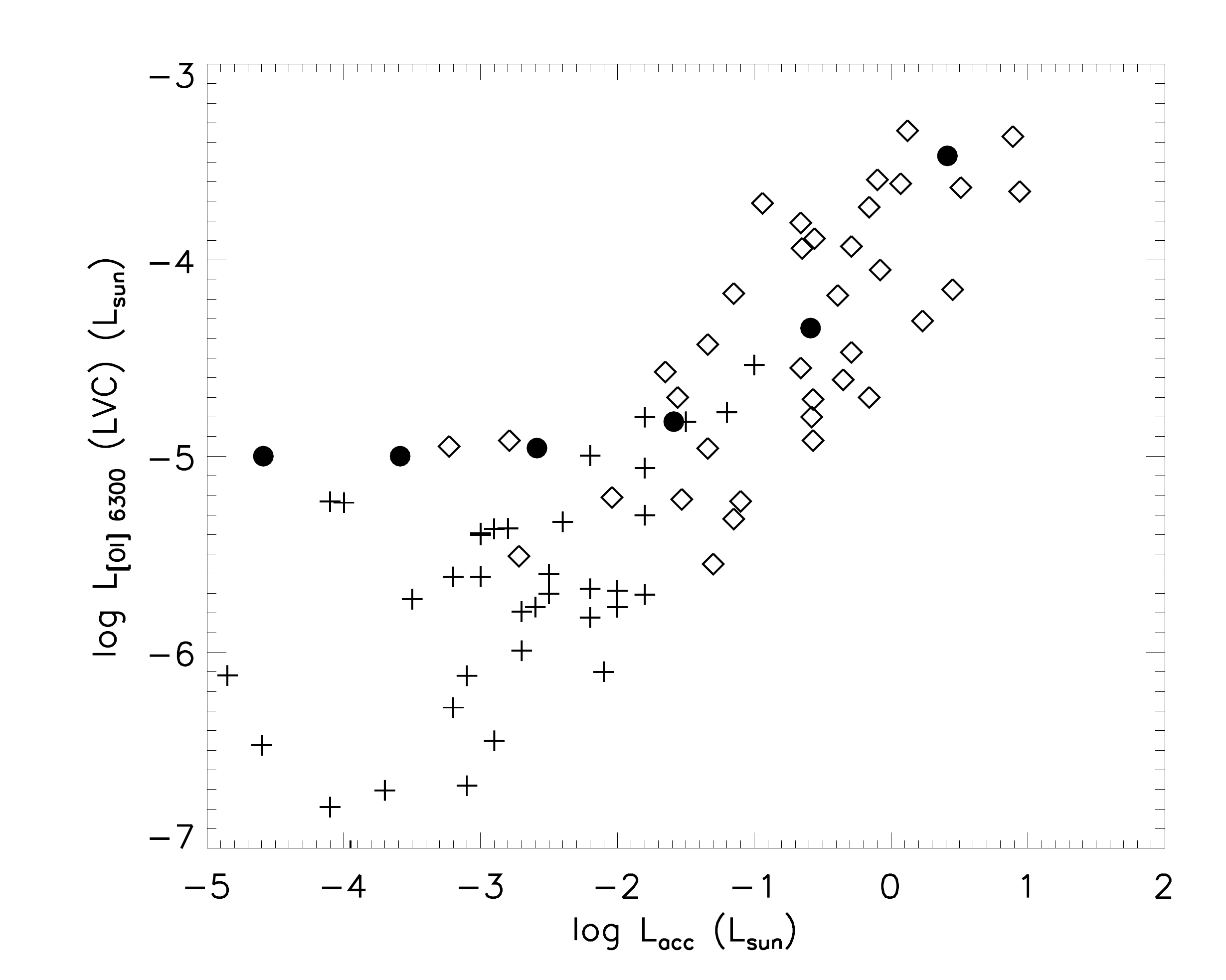}
\caption{[O I] 6300 luminosity versus accretion luminosity for the
Samples I and II objects from Rigliaco et al. (2013) (empty diamonds) and the Lupus sample from Natta et al. (2014) (crosses) compared to the model predictions from this work with constant X-ray luminosity (filled circles). Emission from the X-ray spectrum at $<$100eV is not suppressed in these models. See section 3.2 for details. }
\end{center}
\end{figure}

\subsection{Models at constant X-ray luminosity: correlation with L$_{acc}$}
The observations of Rigliaco et al. (2013) showed a clear relation between the  L$_{[OI]}$ with the FUV luminosity L$_{FUV}$. We suggest that the observed correlation is explainable in terms of the correlation between the emission region of the collisionally excited [OI] line at 6300A and the EUV flux reaching the wind. As the EUV luminosity is dominated by the accretion luminosity of the YSO, we show this point here by running a set of models of constant X-ray luminosity (L$_X = 2\times 10^{30}$erg/sec) and vary only the accretion luminosity, L$_{acc}$ = a $\times$ L$_{bol}$, over a range of  a from 1 to 10$^{-5}$. The results are shown in Figure 2, where the filled circles show our models and the empty symbols the observational results of Rigliaco et al (2013). It is clear from the figure that the observed correlation is reproduced for a {\it constant} X-ray luminosity, and by varying only the input accretion luminosity. 

Both the Rigliaco et al. (2013) observations and the models seem to show that the correlation flattens as the accretion luminosity (i.e. the UV flux reaching the wind) decreases. In the models this is due to the [OI] 6300 line luminosity reaching the floor value set by the X-ray illumination. 

However, Figure 2 also includes the observational sample in the Lupus star forming region from Natta et al. (2014), represented by the crosses. These data extends to much lower accretion luminosities and includes mostly late M stars with masses typically around 0.2~M$_{\odot}$, which is the median of their sample. The Lupus data does not show a flattening in the correlation. It is important to note at this point that our models are not appropriate for a comparison at such low masses, as they were computed for a stellar mass of 0.7~M$_{\odot}$ (note that only 7 out of 44 stars in the Natta et al., 2014, sample have masses larger than 0.5~M$_{\odot}$).  The X-ray properties of such low mass stars are also not very well known. A recent attempt at characterisation of the X-ray properties in the TW Hya region by Kastner et al. (2016), found that log (L$_X$/L$_{\sc bol}$) decreases for stars with spectral type M4 or later, and its distribution broadens. This study, while suffering from poor statistics for stars later than M3, also showed that the later type stars had more long-lived discs, probably due to their less efficient X-ray irradiation. A new modeling campaign to cover this new parameter space is under way and will be the focus of a future work.

\begin{figure}
\begin{center}
\includegraphics[width=0.47\textwidth]{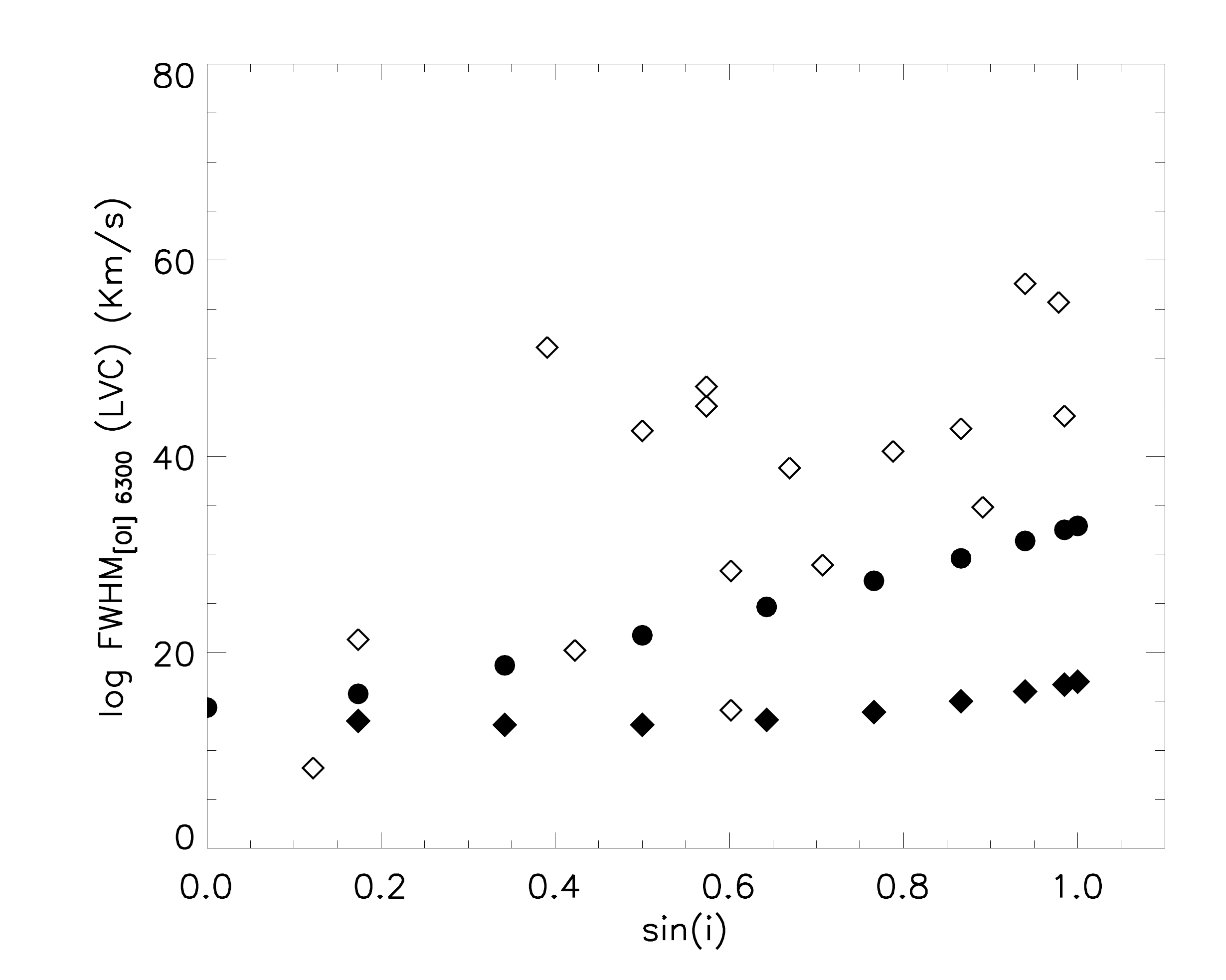}
\caption{FWHM versus sine of the disk inclination from Rigliaco et al. (2013) (empty diamonds) compared to the model predictions for the L$_X $= $2 \times 10^{30}$ erg/sec model of EO10 (filled diamonds) and those from the L$_X $= $2 \times 10^{30}$ erg/sec, L$_{acc}$ = L$_{bol}$ model from this work (filled circles). All model results shown here are for a spectral resolution of R=25000. See section 3.3 for details.}  
\end{center}
\end{figure}

\begin{figure*}
\begin{center}
\includegraphics[width=0.47\textwidth]{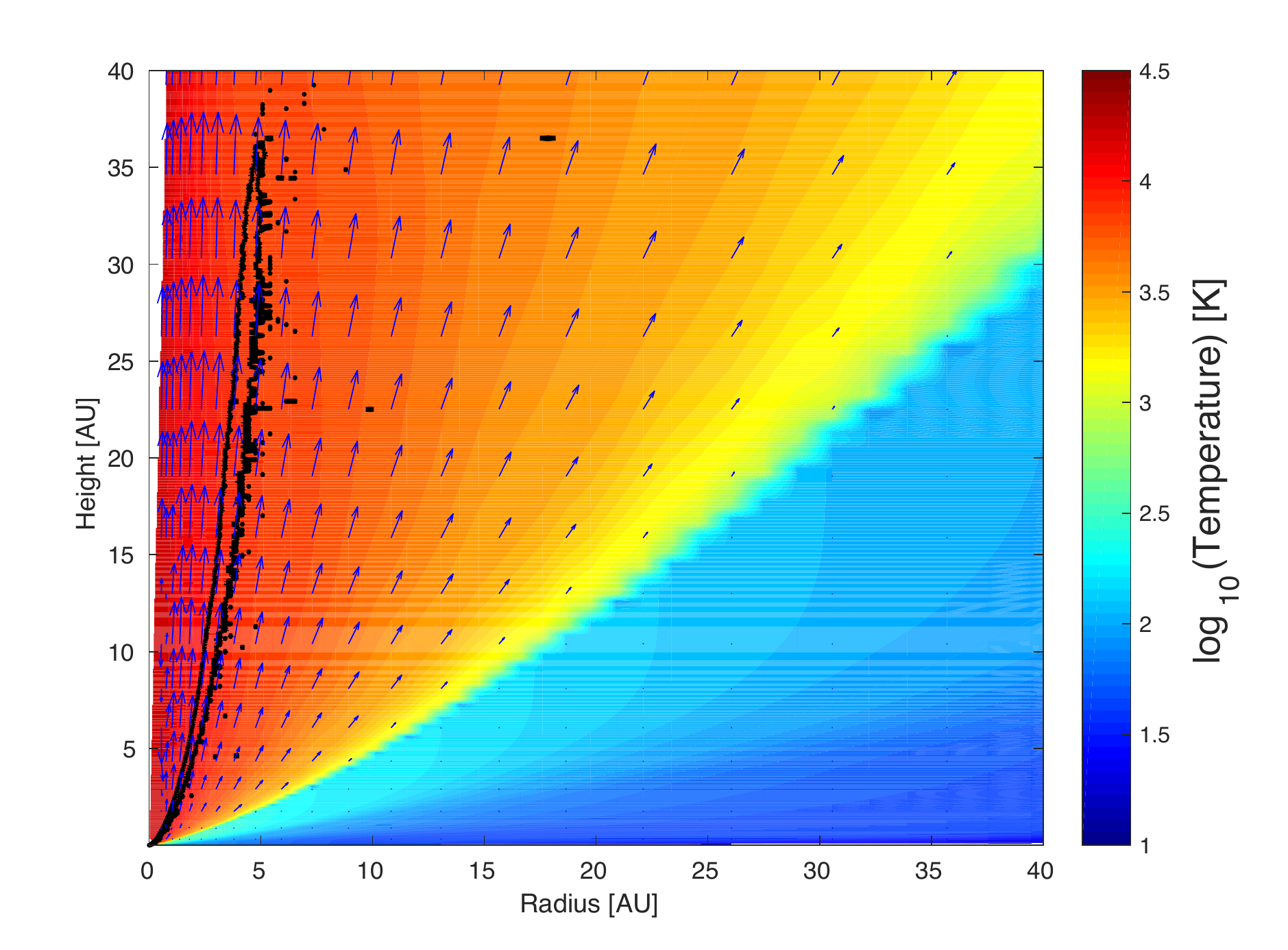}
\includegraphics[width=0.47\textwidth]{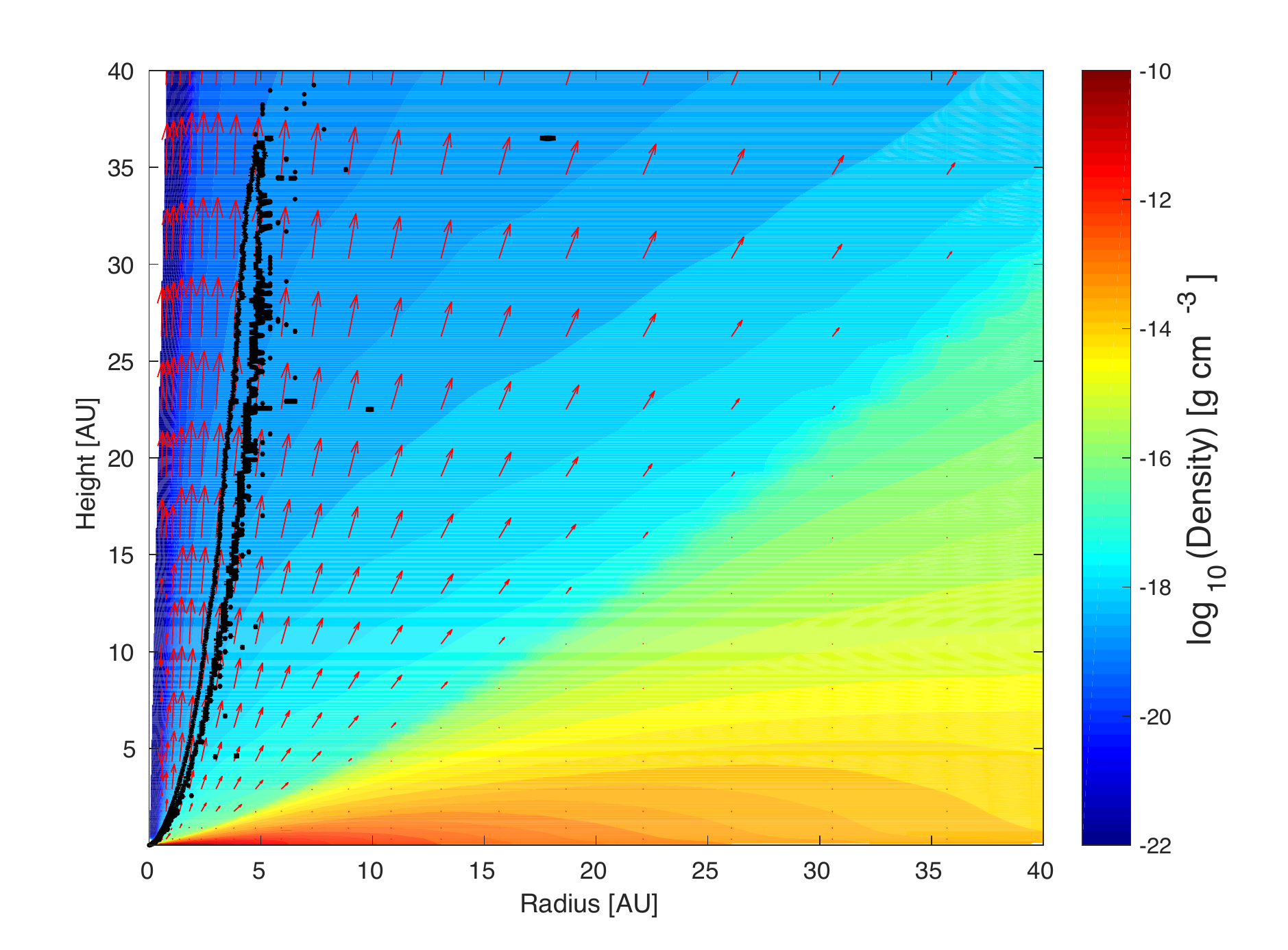}
\caption{Temperature (left) and density (right) maps showing the location of the 85\% emission region of the OI 6300 line (black contour). The velocity field is represented by the red arrows. Plotted is the model with L$_X $= $2 \times 10^{30}$ erg/sec, L$_{acc}$ = L$_{bol}$.}  
\end{center}
\end{figure*}

\subsection{The FWHM of the [OI]6300: a consequence of the emission region}
A further discrepancy between the model predictions of EO10 and the observational results of Rigliaco et al. (2013) are the much lower FWHM of [OI]6300 predicted by the models (filled diamonds in Figure 3) compared to the observations (empty symbols in Fugure 3). As listed in Table 2, the FWHM of the [OI] 6300 line predicted in our new models (e.g. L$_X  =  2\times 10^{30}$ erg/sec and L$_{acc}$ = L$_{bol}$, filled circles in Figure 3) have broader FWHM, which are better in agreement with the observations. The reason for this is that the emission region of the [OI] 6300 line in our new models extends to about 35 AU above the disc (Left panel in Figure 4), compared to only up to 15AU in the models of EO10 (see their Figure 3). The [OI] 6300 line thus samples a wider range of wind velocities, which naturally results in a broader spectral profile. The corresponding line profiles are shown in Figure 6 for 10 inclinations between 0 and 90 degrees and for R=25,000 resolution (similar to the data used by Rigliaco et al. 2013) and R=50,000 (which is more representative of future work). 

We stress that the underlying wind model is the same as that of EO10, the difference in the extension of the [OI]6300 line emission region is purely driven by the presence of the extra accretion luminosity component which is more efficient at heating the wind to the temperatures required to produce the [OI]6300 line in the wind at higher vertical distances above the disc. 

We note however that the FWHM values of the new models are still somewhat narrower than the observational data. The Natta et al. (2014) sample in Lupus (not shown in this plot) have on average still broader profiles, with a median FWHM of 55.5Km/s. New observational data at higher spectral resolution is needed to determine the exact spectral profile of these lines.

\begin{figure}
\begin{center}
\includegraphics[width=0.47\textwidth]{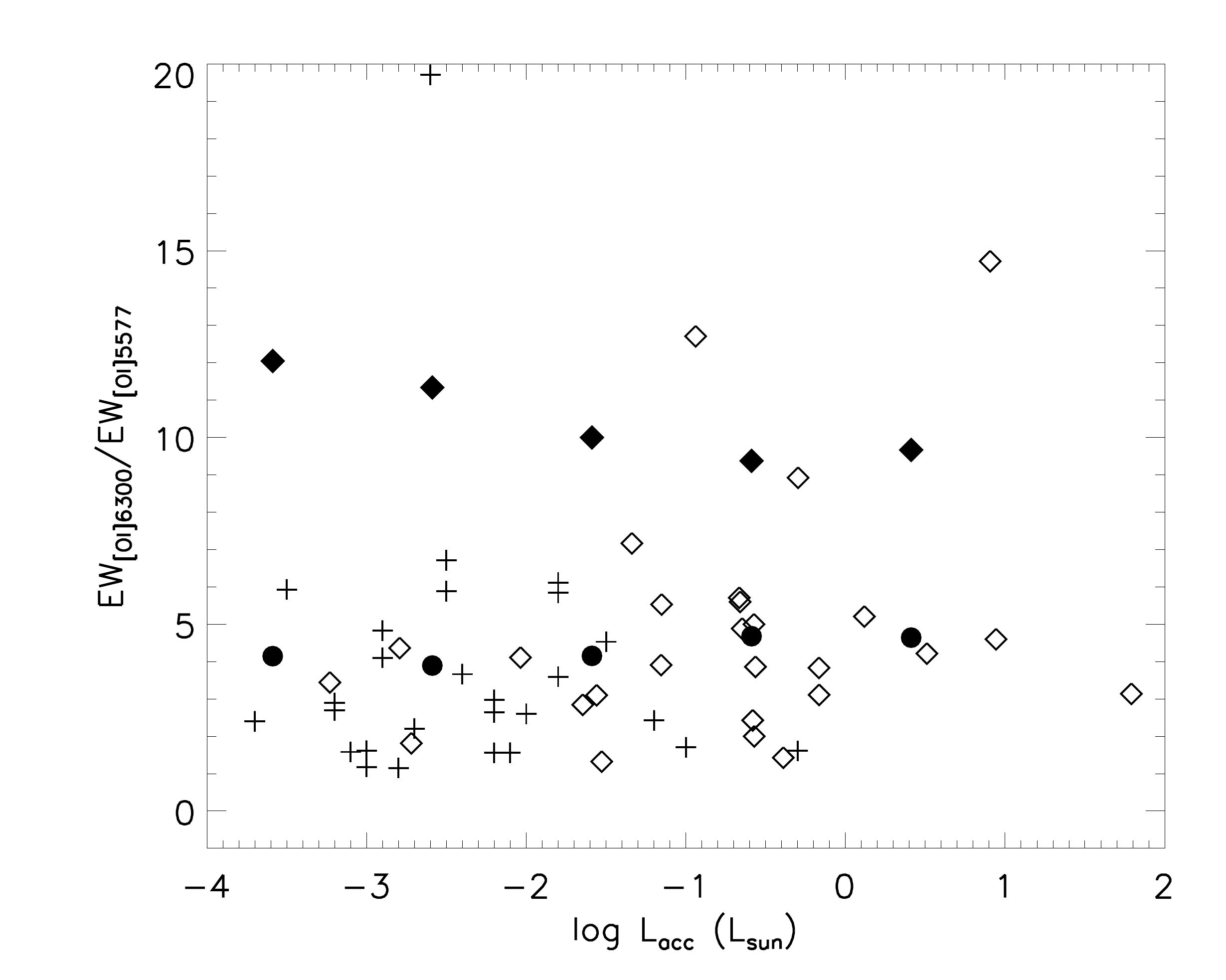}
\caption{EW [OI] 6300A/EW [OI] 5577 A versus accretion luminosity for the observational data of Rigliaco et al. (2013) (empty diamonds) and {Natta et al. (2014) (crosses)}, compared to the model predictions of this work with (filled diamonds) and without (filled circles) neutral hydrogen collision contributions to the [OI] 5577A line. See section 3.4 for details. }
\end{center}
\end{figure}

\subsection{A thermal origin for the [OI] 6300: the  [OI] 6300/ [OI] 5577 ratio}
Perhaps the strongest doubts that the [OI] 6300 may not have a thermal origin were cast by Rigliaco et al (2013) on the basis of their observational mesurements of the [OI]6300/[OI]5577 line ratio in their sample of YSOs. Figure 5 shows that the measurements (empty symbols) scatter around a ratio of about 5, while the models of EO10 (filled diamonds) show significantly higher ratios between 7 and 10, in clear disagreement with the observations. The empty diamonds and the crosses represent, respectively, the data from Rigliaco et al. (2013) and Natta et al. (2014). This discrepancy  led Rigliaco et al. (2013) to favour a non-thermal origin for the formation of the [OI] 6300 line. However, as was already discussed by EO10, the luminosity of the [OI]5577 line calculated from the models is only a lower limit. Both [OI]6300 and [OI]5577 can be excited by collisions with electrons or neutral hydrogen atoms, however in the literature no neutral hydrogen collisional strengths are available for the [OI]5577 line, which is then necessarily underestimated, while the [OI]6300 predictions include both contributions from electron and neutral hydrogen collision. 

To illustrate this point better we have perfomed calculations where we turn off neutral hydrogen collisions for the [OI]6300, so that both [OI]6300 and [OI]5577 only include collisions with electrons. These results are also shown in Figure 5 where the predictions from the electron collision only calculations are shown as filled circles. 
These new calculations are consistent with the observational measurements, implying also that hydrogen collision strengths must contribute in a comparable fashion to the fluxes of both the [OI]6300 and the [OI]5577 lines.

\begin{table}
\begin{tabular}{ccccc}
\hline
L$_X$  &  $<$100eV & L$_{acc}$ &  L$_{[OI]63000}$  & L$_{[OI] 5577}$  \\
$[2E30 erg/sec]$ & suppressed & [L$_{bol}$] &   [L$_{\odot}$] & [L$_{\odot}$] \\
\hline
1.0        & yes & 10      & 2.9e-4 & $>$ 3.0e-5 \\
1.0        & yes & 1         & 3.7e-5 &  $>$3.8e-6 \\
1e-1     & yes & 1         & 1.2e-5 &  $>$2.3e-6 \\
1e-2     & yes & 1         & 1.1e-5 &  $>$2.0e-6 \\

1.0        & yes & 1e-1   & 1.2e-5 & $>$ 7.0e-7 \\
1.0        & yes & 1e-2   & 8.3e-6 & $>$ 3.6e-7 \\
1.0        & yes & 1e-3   & 7.3e-6 & $>$  2.3e-7 \\
1.0        & yes & 1e-4   & 6.2e-6 & $>$ 2.1e-7 \\

1.0        & no  & 10         & 3.4e-4 &  $>$4.e-5 \\
1.0        & no  & 1         & 4.5e-5 &  $>$4.8e-6 \\
1.0       & no   &  1e-1  & 1.5e-5 & $>$1.5e-6 \\
1.0       & no   & 1e-2   & 1.1e-5 &  $>$9.7e-7 \\
1.0       & no   &  1e-3  & 1.0e-5 & $>$8.3e-7 \\
1.0       & no   &  1e-4  & 1.0e-5 & $>$8.2e-7 \\

1.0       &  yes & 1         & $>$2.2e-5& $>$4.7e-6 \\ 
1.0       &  yes  & 1e-1  & $>$5.4e-6 & $>$1.3e-6 \\
1.0       &  yes  & 1e-2  & $>$3.7e-6 & $>$9.5e-7 \\
1.0       &  yes & 1e-3    & $>$3.4e-6 & $>$ 8.2e-7 \\ 
\hline
\end{tabular}
\label{TAB:sum}
\caption{Summary of models and corresponding line luminosity predictions. See text for detail.}
\end{table}

\begin{table}
\centering
\begin{tabular}{ccccc}
\hline
inclination & FWHM$_{25}$ &  v$_{\rm peak}^{25}$ & FWHM$_{50}$ & v$_{\rm peak}^{50}$\\
$[$degrees$]$ & [km~s$^{-1}$] & [km~s$^{-1}$] & [km~s$^{-1}$] & [km~s$^{-1}$] \\
\hline 
0   & 14.36 & -2.1 & 8.97 & -1.6 \\
10 & 15.76 & -2.6 & 11.25 & -2.9\\
20 & 18.67 & -3.6 & 14.82 & -4.4\\
30 & 21.73 & -4.6 & 18.32 & -5.9\\
40 & 24.63 & -5.1 & 21.62 & -6.6\\
50 & 27.29  &-5.1 & 24.63 & -6.9 \\
60 & 29.58 & -4.4 & 27.25 & -6.3\\
70 & 31.36 & -3.4 & 29.33 & -6.1\\
80 & 32.49 & -1.9 & 30.72 & -5.1\\
90 & 32.89 &  0.0 &  31.38 & 0.0\\
\hline
\end{tabular}
\label{TAB:fwhm}
\caption{Velocity at the peak and full-width-half- maximum (FWHM) of the [OI] 6300 line for the L$_X $= $2 \times 10^{30}$ erg/sec, L$_{acc}$ = L$_{bol}$ model from this work as a function of disc inclination. The quantities were calculated for spectral resolutions of 25000 and 50000.}
\end{table}


\begin{figure*}
\begin{center}
\includegraphics[width=0.94\textwidth]{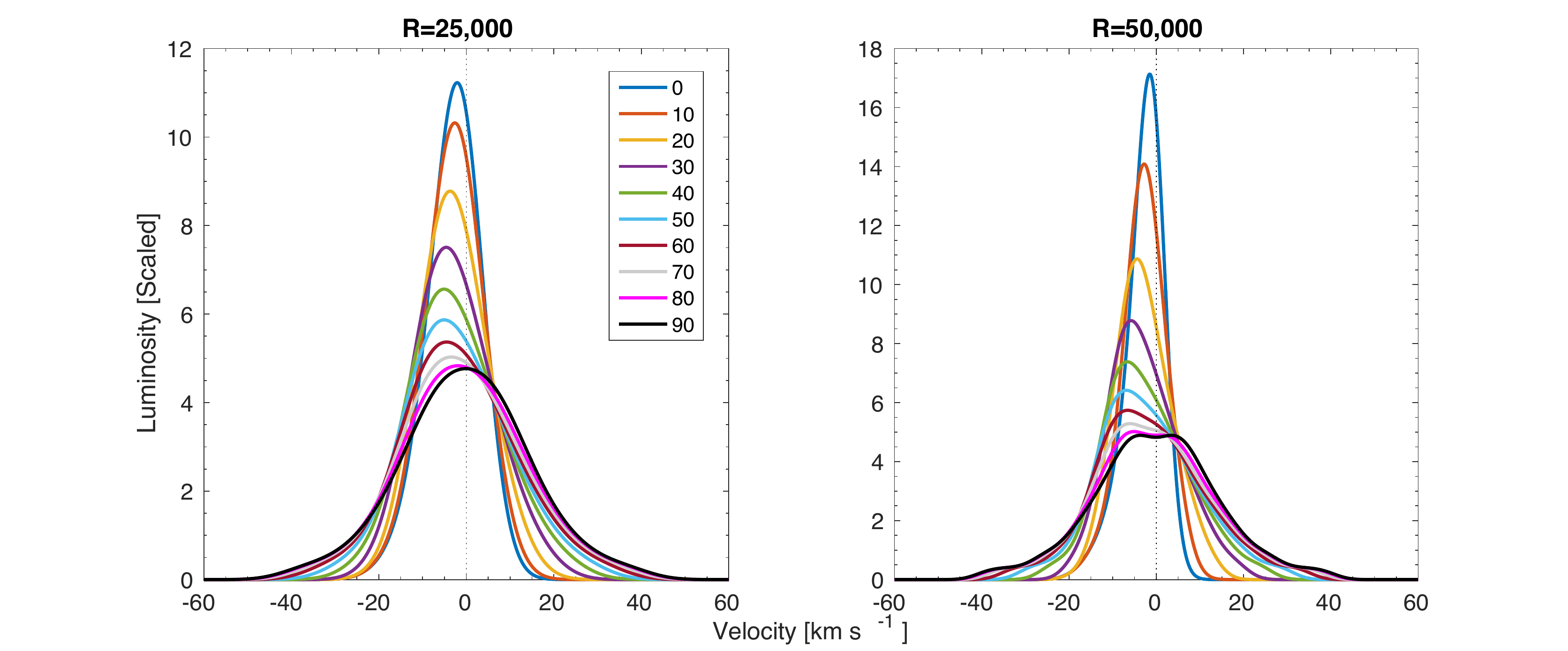}
\caption{[OI] 6300 line profiles from our high resolution hydrodynamical simulations. See text for detailed. The left panel shows line profiles computed with a spectral resolution of 25,000 whereas the right panel is at 50,000. Each line indicates a different viewing inclination, where the inclination is indicated in the legend in degrees.}
\label{fig:profiles}
\end{center}
\end{figure*}

\section{Conclusions}

We have revisited the question of the origin of the low velocity component (LVC) of the blueshifted [OI] line at 6300A. Our new models suggest that this line is produced by collisional excitation of neutral Oxygen by free electrons and neutral hydrogen atoms in the quasi-neutral X-ray photoevaporative winds of young stellar objects up to approximately 1 M$_{\odot}$. Our current models are not relevant to higher mass Herbig-type stars as the photoevaporative wind structure is likely to change due to the fact that these objects have much lower (or absent) X-ray luminosities compared to their solar-like counterparts. 

Our models show that while the wind structure is driven by X-ray radiation (0.1 keV$< E <$1keV) reaching the bound atmosphere of a protoplanetary disc, the size of the emitting region of optical forbidden lines is determined by the EUV (13.6 eV$< E <$100eV) photons reaching the wind. The wind is optically thick to EUV radiation, which thus cannot reach the bound disc atmosphere at all. In fact EUV photons penetrate into and heat only a thin vertically extended (up to about 37 AU) region of the wind above the inner disc ($<$5AU). The luminosity of the [OI] 6300 line and, in fact, of all forbidden lines that have an exponential temperature dependence (due to the Boltzmann term in their emissivity), are strongly weighted to the hottest regions, where neutral hydrogen is still abundant. This is shown in the left panel of Figure 4. This means that the line luminosities scale with the size of the wind region that can be heated by the EUV to the appropriate temperature. From this it follows, as confirmed by our models, that the luminosity of optical forbidden lines is correlated with the accretion luminosity of the YSO. 

As a consequence the luminosity of lines like [OI] 6300 cannot be used to measure wind properties, such as mass loss rate, as their production is not due to the same radiation that causes the photoevaporation of the disc atmosphere. Nevertheless the blueshifted low velocity components detected by (e.g.) Hartigan et al. (1995), Rigliaco et al. (2013) and Natta et al. (2014) are a clear tell-tale sign of a disc wind. High spectral resolution observations of these lines, particularly in combination, remain an important diagnostic tool of mass loss processes from disc atmospheres.

\section{Acknowledgements}
We thank an anonymous referee for helpful comments on the first version of this paper.
We thank Elisabetta Rigliaco for making electronic versions of their data available to us and for the useful discussion.
JEO acknowledges support by NASA through Hubble
Fellowship grant HST-HF2-51346.001-A awarded by the
Space Telescope Science Institute, which is operated by
the Association of Universities for Research in Astronomy, Inc., for NASA, under contract NAS 5-26555.

\label{lastpage}

\end{document}